\documentclass[twocolumn,nofootinbib,amsmath,amssymb,amsfonts,aps,prd,balancelastpage,superscriptaddress]{revtex4-2}

\usepackage{color}
\usepackage{mathtools}
\usepackage[active]{srcltx}
\usepackage{amsthm,amstext,amscd,srcltx}
\usepackage{epsfig,graphicx,bm}
\usepackage{epstopdf, epsf}
\usepackage{dcolumn}
\usepackage{tensor}
\usepackage{graphicx,caption,subcaption}
\usepackage[colorlinks,
           linkcolor = blue,		
           citecolor = blue,
           ]{hyperref}

\newcommand{\be}{\begin{equation}}
\newcommand{\ee}{\end{equation}}

\newcommand{\bse}{\begin{subequations}}
\newcommand{\ese}{\end{subequations}}
\newcommand{\bea}{\begin{eqnarray}}
\newcommand{\eea}{\end{eqnarray}}
\newcommand{\ba}{\begin{array}}
\newcommand{\ea}{\end{array}}
\newcommand{\bc}{\begin{center}}
\newcommand{\ec}{\end{center}}

\begin{document}
\preprint{IPM/P-2012/009}  
\vspace*{3mm}

\title{Sound Speed Resonance of Gravitational Waves in Gauss--Bonnet-coupled inflation}

\author{Andrea Addazi}
\email{addazi@scu.edu.cn}
\affiliation{Center for Theoretical Physics, College of Physics, Sichuan University, 610065 Chengdu, China}
\affiliation{Laboratori Nazionali di Frascati INFN Via Enrico Fermi 54, I-00044 Frascati (Roma), Italy}

\author{Yermek Aldabergenov}
\email{ayermek@fudan.edu.cn}
\affiliation{Department of Physics, Fudan University, 220 Handan Road, Shanghai 200433, China}

 \author{Yifu Cai}
\email{yifucai@ustc.edu.cn}
\affiliation{Deep Space Exploration Laboratory/School of Physical Sciences,
University of Science and Technology of China, Hefei, Anhui 230026, China}
\affiliation{CAS Key Laboratory for Researches in Galaxies and Cosmology/Department of Astronomy,
School of Astronomy and Space Science, University of Science and Technology of China, Hefei, Anhui 230026, China}

\begin{abstract}
\noindent

We demonstrate the occurrence of Sound Speed Resonances (SSR) in Gauss--Bonnet-coupled inflation across a wide range of coupling functions and parameters. After inflation, the damped oscillations of the inflaton around its potential minimum induces damped oscillations of the sound speed of tensor modes, leading to resonant amplification of the latter. Once the inflaton stabilizes around the minimum, the tensor sound speed reduces to unity (speed of light). In the context of multi-field inflation, the sound speed oscillations can be followed by a second phase of inflation, resulting in a distinctive stochastic background of Gravitational Waves (GWs). We show that these GW signals can be probed by upcoming experiments such as {\bf SKA}, {\bf DECIGO}, and {\bf BBO} depending on the duration of the second inflationary phase.

\end{abstract}

\maketitle

\section{Introduction}

The detection of gravitational waves (GWs) from binary compact object mergers \cite{LIGOScientific:2016aoc,LIGOScientific:2017vwq} has unveiled new opportunities for investigating stochastic GW backgrounds arising from new physical mechanisms in the early Universe. GW physics now promises to serve as a valuable tool, alongside large-scale structure (LSS) and the cosmic microwave background (CMB), for testing fundamental cosmological theories.
There are various potential sources for cosmological stochastic GW backgrounds, among which the secondary GW spectrum induced by primordial density perturbations in the early universe stands out and has garnered significant interest recently (refer to Ref.\cite{Domenech:2021ztg} for a comprehensive review). In cosmological perturbation theory, it is established that scalar and tensor perturbations evolve independently at the linear order but become dynamically coupled at the second and higher orders \cite{Acquaviva:2002ud,Baumann:2009ds}. As a result, the scalar mode linked to primordial density perturbations can excite the tensor modes, leading to the generation of secondary GWs. This occurs either when the perturbations are localized at scales much smaller than the Hubble horizon during the inflationary epoch \cite{Biagetti:2013kwa,Fumagalli:2021mpc,Peng:2021zon}, or when they re-enter the horizon during the post-inflationary radiation-dominated (RD) and matter-dominated (MD) eras \cite{Ananda:2006af,Baumann:2007zm,Kohri:2018awv}. This process can produce a relic, detectable stochastic GW background. While  LSS and CMB observations indicate that primordial fluctuations follow an approximately Gaussian profile at scales larger than the Mpc, with a nearly scale-invariant power spectrum and a too low amplitude for current detectors, constraints at much smaller scales ($\ll 1$ Mpc) are less stringent. This allows for the possibility of a significant enhancement of scalar and tensor fluctuations.

Among the possible mechanisms for an amplification of primordial density perturbations, and induced GWs at small scale, the sound speed resonance (SSR) was suggested \cite{Cai:2018tuh,Cai:2020ovp}. 
In case of multi-SSR, it was also shown that chaotic GW resonant spectra are obtained \cite{Addazi:2022ukh}.
These analysis have shown how SSR (of scalar modes) provide a remarkably efficient mechanism for Primordial Black Holes (PBH) production correlated to GW signals. Nevertheless, these analysis were purely phenomenological based on effective parametrization of a modified Mukhanov-Sasaki equation with a time dependent sound speed. In other words, these studies did not seek to explore a potential origin of SSR rooted in a specific theory derived from a Lagrangian principle.

In this paper, we put forward a specific class of models  dynamically realizing the SSR mechanism. We show how SSR is obtained in case of inflaton coupled to a Gauss--Bonnet (GB) term. GB-coupled theories were extensively studied in Refs. \cite{Satoh:2007gn,Satoh:2008ck,Guo:2009uk,Guo:2010jr,Jiang:2013gza,Koh:2014bka,Kawai:2017kqt,Yi:2018gse,Chakraborty:2018scm,Odintsov:2018zhw,Odintsov:2019clh,Rashidi:2020wwg,Kawai:2021bye,Kawai:2021edk,Kawaguchi:2022nku,TerenteDiaz:2023iqk,Easson:2021amu,Odintsov:2022zrj,Elizalde:2023rds,Nojiri:2023mbo}
as a viable 
candidate for inflation or dark energy. Here, we will consider a simple set of models which can be naturally implemented in the inflationary $\alpha$-attractor paradigm inspired by Refs. \cite{Galante:2014ifa,Kallosh:2015lwa,Kallosh:2022feu} (although our mechanism is not limited to $\alpha$-attractors). We will also examine the case of double inflation with the first inflaton coupled to the GB term. In this case, we will demonstrate how  characteristic GW signals by SSRs are obtained. 
These can be tested in future pulsar-timing radio-astronomy experiments or space-based interferometers.

Unless otherwise stated, we work in Planck units, $M_P=1/\sqrt{8\pi G}=1$.

\section{Main setup}
Let us consider an inflaton scalar field coupled to Gauss--Bonnet term,
\begin{equation}\label{L_master}
\sqrt{-g}^{\,-1}\mathcal{L} = \tfrac{1}{2}R-\tfrac{1}{2}(\partial \phi)^2-\tfrac{1}{8}\xi(\phi) R_{GB}^{2}-V(\phi)
\,,
\end{equation}
where $g={\rm det}g_{\mu\nu}$
with $g_{\mu\nu}$ metric tensor, $\phi$ is the (real) inflaton with the potential $V$ and GB coupling $\xi$. The Gauss--Bonnet term defined as 
\begin{equation}
\label{GB}
R_{GB}^{2}\equiv R_{\mu\nu\rho\sigma}R^{\mu\nu\rho\sigma}-4R_{\mu\nu}R^{\mu\nu}+R^{2}\,. 
\end{equation}

In this case, we obtain the following background equations in FLRW, $g_{\mu\nu}={\rm diag}(-1,a,a,a)$:
\begin{align}\label{EoM_phi}
    \ddot{\phi}+3H\dot{\phi}+3\xi_\phi H^2(H^2+\dot{H})+V_\phi &=0~,\\
    \begin{split}\label{EOM_Hdot}
    \dot H(1-\xi_\phi H\dot\phi)+\tfrac{1}{2}\dot\phi^2(1-\xi_{\phi\phi}H^2)\\
    \hspace{2.1cm}-\tfrac{1}{2}\xi_\phi H^2(\ddot\phi-H\dot\phi) &=0~,
    \end{split}\\
    3H^2(1-\xi_\phi H\dot\phi)-\tfrac{1}{2}\dot\phi^{2}-V &=0~,\label{EOM_H}
\end{align}
where subscript $\phi$ means derivative with respect to $\phi$. In order to obtain standard slow-roll regime during inflation, we introduce the following slow-roll parameters
\begin{gather}
\begin{gathered}\label{SR_parameters}
    \epsilon\equiv -\frac{\dot H}{H^2}~,~~~\eta\equiv\frac{\dot\epsilon}{H\epsilon}~,~~~\gamma\equiv\frac{\ddot\phi}{H\dot\phi}~,\\
    \omega\equiv\xi_\phi H\dot\phi~,~~~\sigma\equiv\frac{\dot\omega}{H\omega}~,
\end{gathered}
\end{gather}
and we demand that they are much smaller than one, at least during early stages of inflation.

\section{Scalar modes}

Scalar and tensor perturbations in GB-coupled inflation were discussed in \cite{Hwang:1999gf,Cartier:2001is,Hwang:2005hb,Guo:2009uk,Guo:2010jr,Koh:2014bka,Yi:2018gse,Rashidi:2020wwg,Kawai:2021bye,Kawai:2021edk,Kawaguchi:2022nku,Kawai:2023nqs}. Let us first address scalar perturbations in the presence of the GB coupling. 
Their EOM can be written as
\begin{equation}\label{Scalar_u_mode_eq}
    u''+\Big(C_s^2k^2-\frac{Z_s''}{Z_s}\Big)u=0~,
\end{equation}
where the prime means the derivative w.r.t. the conformal time $\tau$, $k$ is the wave-number and
\begin{align}
    Z_s^2 &\equiv \frac{2a^2(1-\omega)^2}{(1-\tfrac{3}{2}\omega)^2}\bigg[\epsilon-\frac{\omega}{2}(1+\epsilon-\sigma)+\frac{3\omega^2}{4-4\omega}\bigg]~,\label{Z_s_def}\\
    C_s^2 &\equiv 1-2a^2\omega^2\,\frac{\epsilon+\frac{1}{4}\omega(1-5\epsilon-\sigma)}{Z_s^2(1-\tfrac{3}{2}\omega)^2}~.\label{C_s_def}
\end{align}
We will refer to $C_s$ and an analogous function $C_t$ for tensor modes (defined in the next section) as the respective ``sound speeds". Deep inside the horizon, $C_s^2k^2\gg aH$, scalar modes can be described by the usual
Bunch--Davies solution,
\begin{equation}\label{BD_scalar}
    u\simeq \frac{e^{-ik\tau}}{\sqrt{2k}}~,
\end{equation}
provided that the slow-roll conditions for the parameters \eqref{SR_parameters} hold, in which case we have $C_s^2\simeq 1$ and $Z_s''/Z_s\simeq 2a^2H^2$. Once \eqref{Scalar_u_mode_eq} is solved for $u$, it can be related to the comoving curvature perturbation $\zeta$ as $\zeta=u/Z_s$, whose power spectrum is given by
\begin{equation}\label{P_zeta}
    P_\zeta=\frac{k^3}{2\pi^2}|\zeta|^2=\frac{k^3}{2\pi^2}\Big |\frac{u}{Z_s}\Big |^2~.
\end{equation}

Since we are interested in the behavior of the perturbations during oscillations of the inflaton after slow-roll, Eq. \eqref{Scalar_u_mode_eq} is not suitable for numerical solution as it becomes singular whenever $Z_s$ vanishes. This happens periodically during oscillatory stage, which is a known issue in the usual (non-GB) inflationary models as well \cite{Kodama:1996jh,Nambu:1996gf,Finelli:1998bu,Algan:2015xca}. Namely, in the absence of the GB coupling we have $Z_s=\sqrt{2\epsilon}\,a$, and $\epsilon=\dot\phi^2/(2H^2)$, which means that Eq. \eqref{Scalar_u_mode_eq} becomes singular whenever $\dot\phi=0$ (while $Z_s''\neq 0$). In the presence of the GB coupling, $\dot\phi=0$ still leads to vanishing $Z_s$ and the resulting singularity. The vanishing of $Z_s$ can be seen from Eq. \eqref{Z_s_def} by using the definitions of $\omega$ and $\sigma$ from \eqref{SR_parameters}, and the constraint equation \eqref{EOM_Hdot} written in the form
\begin{equation}
    \dot\phi^2/H^2=2\epsilon-\omega(1+\epsilon-\sigma)~.
\end{equation}
Then, in the limit $\dot\phi\rightarrow 0$, we get
\begin{equation}
    Z_s^2\simeq a^2(1+\tfrac{3}{2}\xi^2_\phi H^4)\dot\phi^2/H^2~.
\end{equation}
Therefore, $Z_s$ vanishes when $\dot\phi=0$, and the mode equation \eqref{Scalar_u_mode_eq} encounters singularities during $\phi$-oscillations.

It is worth mentioning that the sound speed $C_s$ remains regular at vanishing $\dot\phi$, although this is not immediately clear from its expression \eqref{C_s_def}. To show this, we again use the explicit forms of the slow-roll parameters \eqref{SR_parameters}, and in the limit $\dot\phi\rightarrow 0$ we find
\begin{equation}
    C_s^2\simeq 1+\frac{2\xi^2_\phi H^2}{1+3\xi^2_\phi H^4/2}(\dot H+\tfrac{1}{4}\xi_\phi H^2\ddot\phi)~.
\end{equation}

In order to avoid the singularities in the evolution of perturbations during oscillations of the inflaton, one can use the gauge-invariant Mukhanov variable \cite{Mukhanov:1988jd}
\begin{equation}\label{Q_def}
    Q\equiv \delta\phi+\frac{\dot\phi}{H}\Psi~,
\end{equation}
as noticed in Refs. \cite{Kodama:1996jh,Nambu:1996gf}. Here $\delta\phi$ and $\Psi$ are gauge-invariant inflaton and spatial metric perturbation (see Eq. \eqref{ds_phi_psi}), respectively. Following Ref. \cite{Finelli:1998bu}, we work in longitudinal gauge where the metric takes the form,
\begin{equation}\label{ds_phi_psi}
    ds^2=-(1+2\Phi)dt^2+a^2(1-2\Psi)dx^2~,
\end{equation}
while the gauge-invariant perturbation $\delta\phi$ from \eqref{Q_def} coincides with the usual inflaton perturbation, so we denote both as $\delta\phi$. We then obtain a (linearized) coupled system of equations for $\Psi$ and $\delta\phi$,
\begin{widetext}
\begin{align}
\begin{split}\label{psi_eq}
    \ddot\Psi &+\Big(3+\frac{\epsilon\omega-\ddot\xi}{1-\omega}\Big)H\dot\Psi+\frac{1-3\omega/2}{1-\omega}H\dot\Phi+\Big[\frac{V}{H^2}-2\omega(1-\epsilon)-\ddot\xi\Big]\frac{H^2\Phi}{1-\omega}+\frac{\xi_\phi H^2}{2(1-\omega)}\delta\ddot\phi\\
    &-\Big[\frac{\dot\phi}{2}-\xi_\phi H^3(1-\epsilon)-\xi_{\phi\phi}H^2\dot\phi\Big]\frac{\delta\dot\phi}{1-\omega}+\Big[\frac{V_\phi}{H^2}+2\xi_{\phi\phi}H\dot\phi(1-\epsilon)+\partial_t(\xi_{\phi\phi}\dot\phi)\Big]\frac{H^2\delta\phi}{2(1-\omega)}=0~,
\end{split}\\[5pt]
\begin{split}\label{delta_phi_eq}
    \delta\ddot\phi &+3H\delta\dot\phi+\Big[\frac{k^2}{a^2}+V_{\phi\phi}+3\xi_{\phi\phi}H^4(1-\epsilon)\Big]\delta\phi-3\xi_\phi H^2\ddot\Psi-3\big[\dot\phi+2\xi_\phi H^3(2-\epsilon)\big]\dot\Psi\\
    &-2\xi_\phi H^2(1-\epsilon)\frac{k^2}{a^2}\Psi-(\dot\phi+3\xi_\phi H^3)\dot\Phi+\Big[2V_\phi+\xi_\phi H^4\Big(\frac{k^2}{a^2H^2}-6+6\epsilon\Big)\Big]\Phi=0~,
\end{split}
\end{align}
\end{widetext}
where the first equation comes from the $ii$ (trace) part of Einstein equations, and the second one from the Klein--Gordon equation. The function $\ddot\xi$ can be written in terms of the slow-roll parameters as $\ddot\xi=\epsilon\omega+\omega\sigma$. 

The $00$ and $0i$ Einstein equations yield the energy constraint
\begin{align}
\begin{split}\label{psi_EC}
&3(1-\tfrac{3}{2}\omega)H\dot\Psi+(1-\omega)\frac{k^2}{a^2}\Psi\\
&+\Big(\frac{V}{H^2}-3\omega\Big)H^2\Phi+\tfrac{1}{2}(\dot\phi+3\xi_\phi H^3)\delta\dot\phi\\
&+\tfrac{1}{2}\Big(V_\phi+\xi_\phi H^2\frac{k^2}{a^2}+3\xi_{\phi\phi}H^3\dot\phi\Big)\delta\phi=0~,
\end{split}
\end{align}
and the momentum constraint equation
\begin{align}
\begin{split}\label{psi_MC}
    &(1-\omega)\dot\Psi+(1-\tfrac{3}{2}\omega)H\Phi+\tfrac{1}{2}\xi_\phi H^2\delta\dot\phi\\
    &-\tfrac{1}{2}(\dot\phi+\xi_\phi H^3-\xi_{\phi\phi}H^2\dot\phi)\delta\phi=0~.
\end{split}
\end{align}
Off-diagonal ($i\neq j$) components of the Einstein equations yield
\begin{equation}\label{psi-phi_eq}
    \Psi-\Phi=-\omega\Phi+\ddot\xi\Psi-\xi_\phi H^2(1-\epsilon)\delta\phi~,
\end{equation}
where the right-hand side (RHS) is the effective anisotropic stress \cite{Saltas:2011xlz} induced by the Gauss--Bonnet coupling. Equation \eqref{psi-phi_eq} can be used to eliminate $\Phi$ as
\begin{equation}
    \Phi=\frac{1-\ddot\xi}{1-\omega}\Psi+\xi_\phi H^2\frac{1-\epsilon}{1-\omega}\delta\phi~.
\end{equation}
Here one might suspect a singularity after the end of slow-roll regime if $\omega=1$. However, it can be shown that $1-\omega$ is positive definite if the scalar potential satisfies $V\geq 0$. This can be seen from the Hubble constraint \eqref{EOM_H},
\begin{equation}
    3H^2(1-\omega)-\tfrac{1}{2}\dot\phi^2=V~,
\end{equation}
which implies that $1-\omega\leq 0$ leads to negative scalar potential. In particular, $\omega=1$ means $\dot\phi\neq 0$ (since $\omega\equiv\xi_\phi H\dot\phi$), which in turn implies negative scalar potential. It is also evident that $\dot\phi=0$ singularity is avoided in Eqs. \eqref{psi_eq}--\eqref{psi-phi_eq} (of course, when deriving the power spectrum of $\zeta$ \eqref{P_zeta}, the singularities will still appear due to the division by $Z_s$).

Next, Eqs. \eqref{psi_eq} and \eqref{delta_phi_eq} are usually combined into a single equation for the Mukhanov variable $Q$ by using \eqref{Q_def} and the constraint equations \cite{Kodama:1996jh,Nambu:1996gf}. However, in the presence of the GB coupling the resulting equation becomes unwieldy, and therefore we will solve the coupled system \eqref{psi_eq} and \eqref{delta_phi_eq} directly, and combine the solutions to obtain $Q$.

The initial conditions for $\delta\phi$ and $\Psi$ can be found as follows. Combining Eqs. \eqref{psi_eq}, \eqref{delta_phi_eq}, and the constraints \eqref{psi_EC}, \eqref{psi_MC}, in slow-roll, subhorizon limit, we obtain
\begin{equation}\label{delta_phi_BD_eq}
    \delta\ddot\phi+3H\delta\dot\phi+\frac{k^2}{a^2}\delta\phi\simeq 0~,
\end{equation}
which has the usual form since the Gauss--Bonnet contribution is suppressed in this regime. For canonically quantized $\delta\phi$, the normalized Bunch--Davies solution of \eqref{delta_phi_BD_eq} reads
\begin{equation}
    \delta\phi_{\rm BD}=\frac{e^{-ik\tau}}{a\sqrt{2k}}~.
\end{equation}
From the constraint Eqs. \eqref{psi_EC} and \eqref{psi_MC} we also find
\begin{equation}
    \Psi_{\rm BD}\simeq\frac{a^2H\dot\phi}{2k^2}\big(1-\frac{\omega}{2\epsilon-\omega}\frac{k^2}{a^2H^2}+\frac{ik}{aH}\big)\delta\phi_{\rm BD}~.
\end{equation}
At the leading order in slow-roll parameters and $aH/k$, from \eqref{Q_def} we have $Q_{\rm BD}\simeq\delta\phi_{\rm BD}$, which is consistent with the BD solution \eqref{BD_scalar} found in the comoving (uniform field) gauge. The consistency can be seen by relating $Q$ and $u$ to the curvature perturbation, $\zeta=u/Z_s=HQ/\dot\phi$.

By using $\delta\phi_{\rm BD}$ and $\Psi_{\rm BD}$ as initial conditions, the system \eqref{psi_eq} and \eqref{delta_phi_eq} can be solved numerically, and the solutions can be combined as $\delta\phi+\dot\phi\Psi/H=Q$. The power spectrum in terms of $Q$ reads
\begin{equation}
    P_\zeta=\frac{k^3}{2\pi^2}|\zeta|^2=\frac{k^3H^2}{2\pi^2\dot\phi^2}|Q|^2~.
\end{equation}
As mentioned earlier, although $\zeta$ and $P_\zeta$ have singularities when $\dot\phi=0$, the solution $Q$ is regular.

Our main focus is the tensor modes and their amplification, but nonetheless, it is important to ensure the linearity and PBH non-overproduction by scalar perturbations.

\section{Tensor modes}

Tensor modes do not encounter singularities when $\dot\phi=0$, so the usual treatment can be applied. We define tensor perturbations $h_{ij}$ via the metric
\begin{equation}
    ds^2=-dt^2+(\delta_{ij}+ h_{ij})dx^idx^j~,
\end{equation}
and project their Fourier modes into the chiral basis,
\begin{equation}
    \tilde h_{\lambda}(t,{\bf k})\equiv e_{-\lambda}^{ij}h_{ij}(t,{\bf k})~,
\end{equation}
where $\lambda=(+,-)$, $e^{ij}_\lambda=e^i_\lambda e^j_\lambda$ with the polarization tensor $e^i_\lambda$. The tensor $e^{ij}_\lambda$ satisfies $e^{ij}_{\lambda}e_{ij,\lambda}=0$ and $e^{ij}_{\lambda}e_{ij,-\lambda}=1$, while $e^i_\lambda$ satisfies ${\bf k}\cdot{\bf e}_\lambda=0$, $i{\bf k}\times {\bf e}_\lambda=\lambda k{\bf e}_\lambda$, ${\bf e}^*_\lambda={\bf e}_{-\lambda}$, and ${\bf e}_\lambda({\bf k})={\bf e}_{-\lambda}(-{\bf k})$, where we use boldface letters to denote 3D vectors. We quantize $\tilde h_\lambda(t,{\bf k})$ as
\begin{equation}
    \tilde h_{\lambda}(t,{\bf k})=h_\lambda(t,{\bf k})\hat a_{{\bf k},\lambda}+h^*_\lambda(t,{\bf k})\hat a_{-{\bf k},\lambda}^\dagger~,
\end{equation}
with creation and annihilation operators satisfying
\begin{equation}
    [\hat a_{{\bf k},\lambda},\hat a^{\dagger}_{\bf k',\lambda'}]=\delta_{\lambda,\lambda'}\delta^3({\bf k}-{\bf k'})~,
\end{equation}
while the Wronskian normalization condition reads
\begin{equation}
    h_\pm {h^*}'_\pm-h_\pm'h^*_\pm=4i/a^2~.
\end{equation}

Equation of motion for the mode function $h_\lambda$ can be written in terms of the new variable $y=Z_t h_\lambda$ (same for both helicities)~\footnote{Since there are no parity-violating interactions, we can drop polarization index on $y$ as long as we remember to count both contributions when computing the power spectrum.}:
\begin{equation}\label{y_eq}
    y''+\bigg(C_t^2k^2-\frac{Z_t''}{Z_t}\bigg)y=0~,
\end{equation}
where
\begin{align}\label{C_t_def}
    Z_t^2 &\equiv a^2(1-\omega)~,\\
    C_t^2 &\equiv \frac{1-\omega(\epsilon+\sigma)}{1-\omega}~.
\end{align}
Bunch--Davies initial condition for the tensor mode $y$ is given by $y_{\rm BD}=\sqrt{2/k}e^{-ik\tau}$, provided that the slow-roll parameters are small at this time.

The power spectrum of tensor modes reads,
\begin{equation}
    P_h=\frac{k^3}{2\pi^2}\sum_\lambda|h_\lambda|^2~.
\end{equation}
Another useful quantity is the GW density fraction today,
\begin{equation}\label{Omega_GW}
    \Omega_{\rm GW}h^2\simeq\Omega_{r,0}h^2P_h/24~,
\end{equation}
where $h\approx 0.67$ is dimensionless Hubble parameter, and $\Omega_{r,0}h^2\approx 4.15\times 10^{-5}$ is the radiation density today.

\section{Benchmark model}

As an example, in the Lagrangian \eqref{L_master} we choose a simple $\alpha$-attractor potential \cite{Galante:2014ifa},
\begin{equation}\label{V_alpha_attr}
V=\tfrac{1}{2}M^2\big(1-e^{-\sqrt{\frac{2}{\alpha}}\phi}\big)^2~,
\end{equation}
with parameters $\alpha$ and $M$. Inflation in this case happens at large positive $\phi$. In order to fix Gauss--Bonnet coupling function, we assume that it does not spoil slow-roll inflation during the horizon exit of CMB scales, but becomes important towards the end of inflation, in particular, during the oscillation phase of the inflaton. This can be realized naturally in the exponential $\alpha$-attractor framework if we choose for instance,
\begin{equation}\label{GB_function}
    \xi(\phi)=\lambda e^{-n\sqrt{\frac{2}{\alpha}}\phi}~,
\end{equation}
with some constants $\lambda$~\footnote{Upon restoring the Planck mass, we write the Lagrangian as ${\cal L}\sim \frac{M_P^2}{2}(R+\xi R_{\rm GB}^2)$, so that the parameter $\lambda$ has the dimension [mass]$^{-2}$.} and $n>0$. During early inflation, when $\phi\ll 1$, the GB coupling is suppressed, while towards the end of slow-roll it can be activated for appropriately chosen $\lambda$.

\begin{figure}
\centering
  \centering
  \includegraphics[width=1\linewidth]{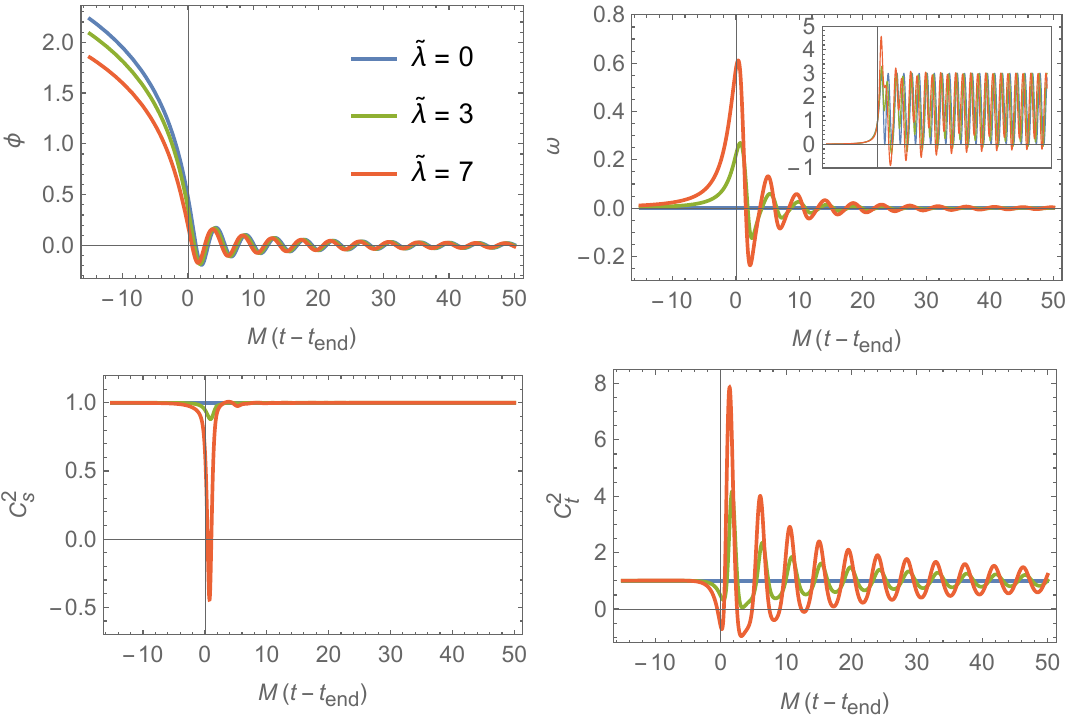}
\captionsetup{width=1\linewidth}
\caption{Top-left: the evolution of $\phi$. Top-right: GB slow-roll parameter $\omega$ (the inset plot shows the Hubble slow-roll parameter $\epsilon$ in the same time interval). Bottom-left: scalar mode sound speed. Bottom-right: tensor mode sound speed. We define $\tilde\lambda\equiv\lambda M^2$.}\label{Fig_plot_set_single}
\end{figure}

By choosing $\alpha=n=1$, in Figure \ref{Fig_plot_set_single} we show the numerical background solution $\phi$ (top-left), GB slow-roll parameter $\omega$ ($\epsilon$ is shown in the inset plot) (top-right), sound speed of scalar modes (bottom-left), and sound speed of tensor modes (bottom-right). The plots show normalized time near the end of inflation, $\epsilon(t_{\rm end})=1$, and $\tilde\lambda\equiv\lambda M^2$ ($\tilde\lambda$ is dimensionless, since in our convention $[\lambda]=[{\rm mass}]^{-2}$). The plots of $C_s$ and $C_t$ show their qualitatively different behaviour: $C_s^2$ has a single dip (with barely visible features afterwards), while $C_t^2$ oscillates with gradually decaying amplitude. This difference can be explained by noticing in \eqref{C_s_def} and \eqref{C_t_def} that the time-dependent part of $C_s^2$ is at least quadratic in $\omega$, i.e. $C_s^2=1+{\cal O}(\omega^2)$, while for $C_t^2$ we have $C_t^2=1+{\cal O}(\omega)$. Since the magnitude of $\omega$ is quickly reduced after initial spike (see Figure \ref{Fig_plot_set_single}), the oscillations of $C_s^2$ are suppressed by $\omega^2$, while the oscillations of $C_t^2$ are suppressed by $\omega$.

For large enough GB coupling, both $C_s^2$ and $C_t^2$ can temporarily become negative, as is shown in Figure \ref{Fig_plot_set_single} for $\tilde\lambda=7$. This leads to an instability, and exponential growth of the perturbations during the time when the corresponding sound speed is imaginary. This instability can lead to interesting consequences during the reheating epoch. For example, in Ref. \cite{Kawai:2017kqt} the authors proposed a leptogenesis scenario via the aforementioned instability of tensor modes when both Gauss--Bonnet and gravitational Chern--Simons terms are present. In \cite{Kawai:2017kqt}, the authors use a specific form of the GB coupling inspired by superstring compactifications. We find that the oscillatory behaviour of $C_t^2$, as well as transient instabilities as shown in Figure \ref{Fig_plot_set_single} can be expected from (but not limited to) a broad class of GB couplings satisfying the requirements mentioned earlier: initial suppression during early inflation, and growth at the end of slow-roll. This class also includes the GB couplings inversely proportional to the scalar potential, $\xi\sim 1/V$ \cite{Guo:2010jr,Jiang:2013gza,Yi:2018gse,Rashidi:2020wwg,ElBourakadi:2021nyb}.

Our focus is on the possibility of sound speed resonance of tensor modes, and its effect on Stochastic Gravitational Wave Background (SGWB), we therefore will avoid the sound speed instabilities by requiring that $C_{s/t}^2>0$. We consider the situation where inflation continues after the transient slow-roll violation and the oscillation of $C_t^2$ shown in Figure \ref{Fig_plot_set_single}, such that these modes can exit the horizon and re-enter some time after the reheating epoch, contributing to (potentially observable) SGWB. This can be achieved, for example, in a double inflation scenario where $\phi$ drives the first stage of inflation, and another scalar $\chi$ drives the second stage, with transient slow-roll violation \cite{Kofman:1985aw,Silk:1986vc,Polarski:1992dq}.

\section{Two-field inflation and observable SGWB}

Specifically, we consider the following simple extension of our benchmark model,
\begin{align}\label{L_two-field}
\begin{aligned}
\sqrt{-g}^{\,-1}\mathcal{L} &= \tfrac{1}{2}[R-(\partial \phi)^2-(\partial \chi)^2]-\tfrac{1}{8}\xi(\phi) R_{GB}^{2}\\
&-\tfrac{1}{2}M^2\big[\big(1-e^{-\sqrt{\frac{2}{\alpha}}\phi}\big)^2+\gamma^2\chi^2\big]\,,
\end{aligned}
\end{align}
where we added a minimally-coupled scalar $\chi$ with quadratic potential which is suppressed, $\gamma\ll 1$, against the $\alpha$-attractor potential. This will ensure that during the first stage, when $\phi$ is slowly rolling, the second scalar $\chi$ is frozen (at some non-zero value). Once $\phi$ reaches its local minimum, slow-roll can continue in the $\chi$-direction, after its transient violation. GB function is given by \eqref{GB_function}.

The resulting numerical plots for background fields are similar to those shown in Figure \ref{Fig_plot_set_single}, with the exception that $t_{\rm end}$ now indicates the end of the first stage, and $\epsilon$ starts to decrease after some oscillations, due to the resumption of slow-roll. 

In order to derive the gravitational wave power spectrum, we numerically solve Eq. \eqref{y_eq}, where the slow-roll parameters and the scale factor are given by the (numerical) background solution of the two-field model \eqref{L_two-field}. We start numerical integration of \eqref{y_eq} deep inside the horizon, e.g. at the time $t_i$ when $C_tk=100aH$, and stop the integration when the mode is well outside the horizon, e.g. at the time $t_f$ when $C_tk=aH/100$. The exception is when $t_i$ and $t_f$ fall in the period of transient slow-roll violation (when $\phi$ starts oscillating around its local minimum), in which case we move $t_i$ well before the slow-roll violation happens (such that the mode is in the Bunch--Davies vacuum), and $t_f$ well after the slow-roll regime is re-established (to allow the solution to stabilize after the oscillatory phase). The value of the solution $y$ is then taken at the end of numerical integration ($t_f$).

\begin{figure}
\centering
  \centering
  \includegraphics[width=1\linewidth]{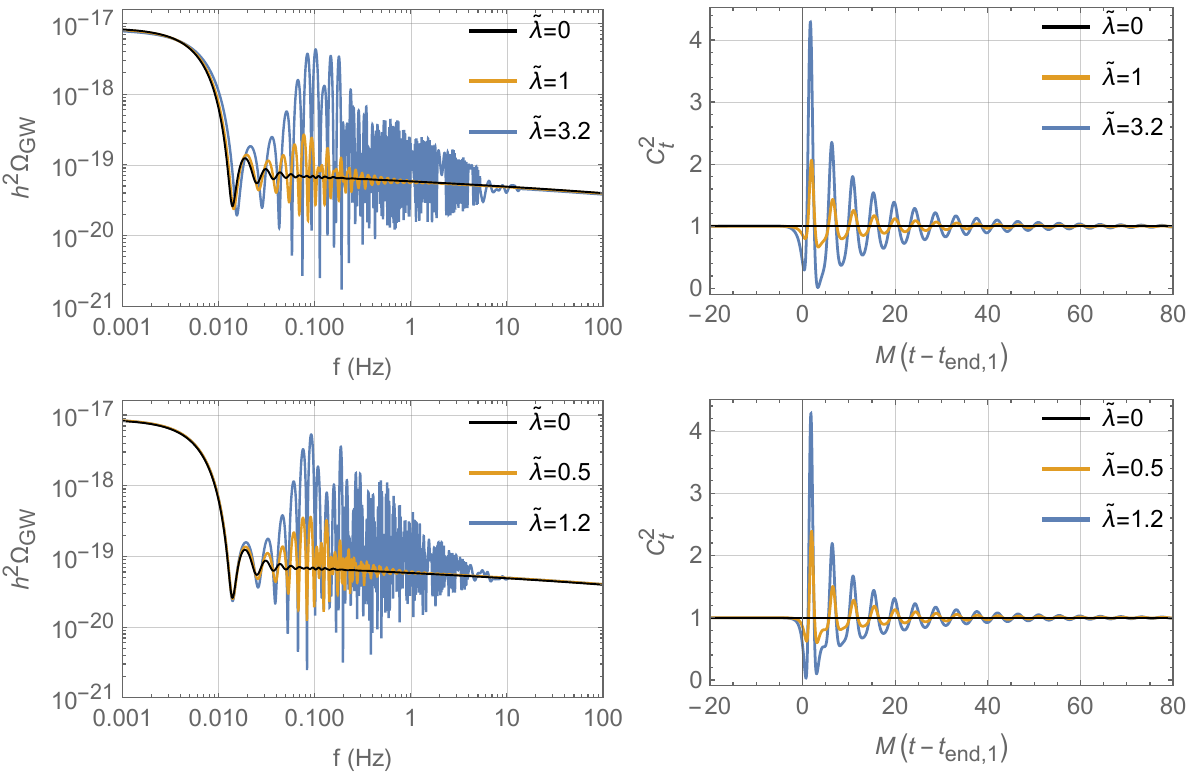}
\captionsetup{width=1\linewidth}
\caption{The GW density fraction and the squared sound speed for the model \eqref{L_two-field} with GB coupling \eqref{GB_function}. First row: $n=1$. Second row: $n=2$.}\label{Fig_SSR_n}
\end{figure}

The resulting GW density fraction is shown in Figure \ref{Fig_SSR_n}, where resonant amplification of tensor modes can be seen for the cases $n=1$ (first row) and $n=2$ (second row), with $\alpha=1$ and $\gamma=0.01$ in both cases (we set the initial value $\chi_0=9$ which results in $\sim 22$ efolds of $\chi$-driven second phase of inflation). The first column shows GW power spectrum in terms of the density fraction $h^2\Omega_{\rm GW}$, and the second column shows the respective squared sound speeds around the end of the first phase of inflation. The inflaton mass parameter is adjusted around $M\sim 1.3\times 10^{-5}$ by CMB normalization of the scalar power spectrum, $P_\zeta(k_{\rm CMB})\approx 2.1\times 10^{-9}$. The squared sound speeds of tensor and scalar modes are positive in both cases of Figure \ref{Fig_SSR_n}. We find taht the scalar perturbations do not experience any amplification in our models and do not contribute to either PBH abundance or secondary GWs.

The growth that is seen in Figure \ref{Fig_SSR_n} is purely due to the oscillations of $C_t^2$ in the mode equation \eqref{y_eq} (since $\chi$ is not coupled to the GB term, the tensor mode equation is unchanged, but the scalar mode equations \eqref{psi_eq} and \eqref{delta_phi_eq} are modified by the presence of $\delta\chi$). As can be seen, the level of amplification of tensor modes grows with $\tilde\lambda$. However, the sound speed plots show that $\tilde\lambda$ is bounded from above if we want to keep $C_t^2>0$. Therefore, the amount of GW amplification by our SSR mechanism is limited for a given model.

Let us comment on the amplified frequencies of GWs. Since the sound speed oscillations last for a time period which is much shorter than the duration of inflation, we can identify the characteristic frequency of the amplified GWs with the time when the first stage of inflation ends. This is when the (largest) oscillations of the sound speed take place. The frequency is related to the wavenumber as $k=2\pi f$, while $k$ can be related to the elapsed number of efolds as $k=aH=\beta He^N$ (at horizon crossing), where $\beta$ is a constant of normalization of $a$. If we take $N$ as the number of efolds from the horizon exit of the CMB scale $k_*$ (i.e. from zero to $\sim 55$), then from $k_*=\beta H_*$ we can find $\beta$. This allows us to express the frequency as
\begin{equation}\label{f_A}
    f=\frac{k_*H}{2\pi H_*}e^N~\Longrightarrow~~f_{A}=\frac{k_*H_{A}}{2\pi H_*}e^{N_A}~,
\end{equation}
where $f_A$ is the amplified frequency, while $H_A$ and $N_A$ are the values when SSR takes place, which can be taken as the end of the first slow-roll phase driven by $\phi$. In the model \eqref{L_two-field} we have $H_*\approx M/\sqrt{6}$, and for an order-of-magnitude estimate of $H_A$, we can approximate it by setting $\phi=\dot\phi=0$, so that $H_A\sim M\gamma\chi_0/\sqrt{6}$, where $\chi_0$ is the initial value of $\chi$ at which it is frozen until the end of the first slow-roll.

Let us apply Eq. \eqref{f_A} to the parameter choice of Figure \ref{Fig_SSR_n}: $\gamma=0.01$ and $\chi_0=9$. In this case, the first slow-roll phase ends at $N\sim 33$, which can be used as the value of $N_A$. In Planck units ($c=\hbar=M_P=1$) we have ${\rm Mpc}\sim 4\times 10^{56}$ so that $k_*=0.05~{\rm Mpc}^{-1}\sim 1.2\times 10^{-58}$. Once $f_A$ is calculated in Planck units, the Hertz unit can be restored by using its value in Planck units, ${\rm Hz}=1/{\rm s}\sim 2.6\times 10^{-43}$. The result is $f_A\sim 0.2~{\rm Hz}$, which is a good estimate for the numerical results of Figure \ref{Fig_SSR_n}. Notice that this frequency estimate does not depend on the Gauss--Bonnet parameters, but on the time of transition from the first to the second slow-roll stage. However, the Gauss--Bonnet parameters determine the peak amplitude of the GWs. Although analytical treatment of the GW SSR effect is difficult in the Gauss--Bonnet-coupled inflation, some insights can be drawn from using analytical modeling of the sound speed function, as is done in Refs. \cite{Cai:2020ovp,Addazi:2022ukh}.

\begin{figure}
\centering
  \centering
  \includegraphics[width=1\linewidth]{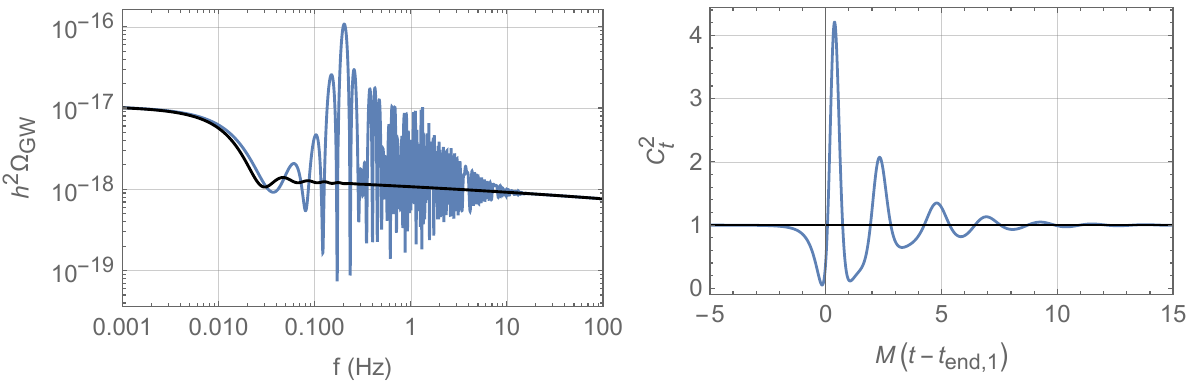}
\captionsetup{width=1\linewidth}
\caption{Resonant amplification of tensor modes for GB coupling \eqref{xi_poly}. Black curve shows the case of no GB coupling compared to the inclusion of it (blue curve).}\label{Fig_SSR_poly}
\end{figure}

In principle, by testing different scalar potentials and GB functions, one can obtain a larger amplification (in fact, slightly increasing $\gamma$ in \eqref{L_two-field} can also lead to somewhat larger peaks, as will be shown below) than that of Figure \ref{Fig_SSR_n}. For example, we find that for
\begin{equation}\label{xi_poly}
    \xi(\phi)=\lambda_1/(\lambda_2+\phi^2)~,
\end{equation}
one can generate a larger peak shown in Figure \ref{Fig_SSR_poly}. The parameter choice for this plot is $\lambda_1 M^2=0.025$, $\lambda_2=0.041$, $\gamma=0.04$, and $\chi_0=11.5$.

\begin{figure}
\centering
  \centering
  \includegraphics[width=1\linewidth]{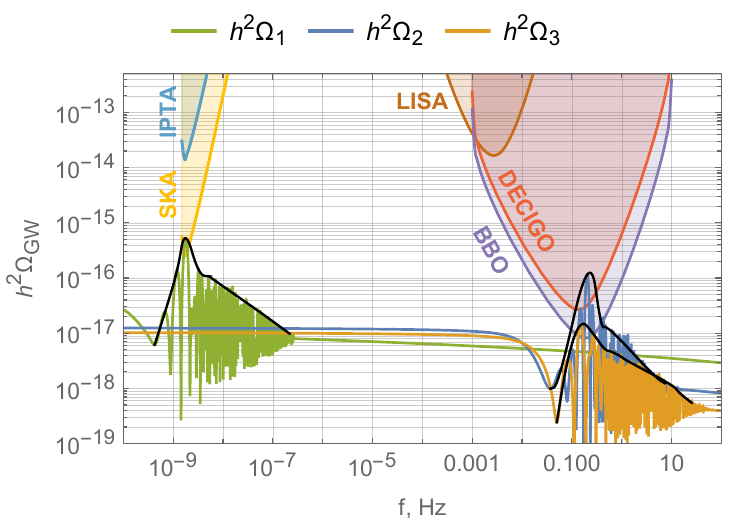}
\captionsetup{width=1\linewidth}
\caption{Three examples of potentially observable GW spectra in our models are displayed in green, blue and orange respectively. See main text for parameter sets. In black, we over-impose results obtained from an envelope approximation. }\label{Fig_Omega}
\end{figure}

Finally, in Figure \ref{Fig_Omega} we plot several examples of SGWB generated by our SSR mechanism, and compare them with sensitivity curves of the future GW experiments, including spaceborne interferometers (LISA \cite{LISA:2017pwj}, DECIGO \cite{Kudoh:2005as}, BBO \cite{Crowder:2005nr}), and pulsar timing array (PTA) experiments (IPTA \cite{Hobbs:2009yy} and SKA \cite{Carilli:2004nx}). We use power-law-integrated curves, see e.g. \cite{Schmitz:2020syl} for details. The shown examples and their parameter sets are as follows. For $\Omega_1$: GB function of \eqref{xi_poly} with $\{\lambda_1M^2=0.025,\lambda_2=0.0348,\gamma=0.04,\chi_0=15.9\}$. For $\Omega_2$: the same GB function with  $\{\lambda_1M^2=0.025,\lambda_2=0.041,\gamma=0.04,\chi_0=11.5\}$. For $\Omega_3$: GB function \eqref{GB_function} with $\{n=1,\lambda M^2=3.8,\gamma=0.03,\chi_0=10\}$ (all models use $\alpha=1$).

Figure \ref{Fig_Omega} shows that the GB-induced SSR in certain cases can be tested by more sensitive space-borne detectors, like BBO and DECIGO, as well as the SKA  radio telescope project.

\section{Conclusions}

In this paper, we have shown how the Sound Speed Resonance (SSR) phenomenon can be realised in Gauss--Bonnet-coupled inflation in a wide parametric space. 
Although we focus on a simple class of models 
inspired by inflationary $\alpha$-attractors, such SSR can be realized in any inflationary model for a suitably chosen GB-coupling, at least in principle.

First, we have demonstrated a time-variation of the sound speeds of scalar and tensor modes driven by the Lagrangian interactions, by focusing on the regime of damped oscillations of the inflaton around its minimum. These oscillations lead to damped oscillations of the tensor mode sound speed, while the variations of the scalar mode sound speed are suppressed by comparison. Then, we showed how the GB-induced SSR can generate a large stochastic background of GWs if the oscillatory phase is followed by more e-folds, as can be realized in two-field double inflation scenario. Notably, we obtained several examples of GW signals which can be tested in future experiments in radio astronomy and space-based interferometers \footnote{An alternative GW phenomenology of (single) GB-coupled inflation, in the frontier of ultra-high-frequencies (UHF) far beyond LIGO,  was recently studied in Refs. \cite{Mudrunka:2023wxy,Tokareva:2023mrt}. Moreover UHF GWs can also have an important impact 
in radio astronomy from stochastic resonant graviton-photon transitions with cosmological magnetic fields \cite{Addazi:2024osi,Addazi:2024kbq}.}. This represents an important step towards the 
embedding of the SSR phenomenon, for GWs and PBHs co-genesis, in a theory of inflation based on a Lagrangian principle and fundamental symmetries. 
Several results show how inflationary $\alpha$-attractors can be obtained from models based on superconformal groups, supergravity
and string theory \cite{Ferrara:2013rsa,Kallosh:2013yoa,Cecotti:2014ipa}, while Gauss--Bonnet portals can be generated by string theory corrections \cite{Antoniadis:1993jc,Kawai:1998ab,Kawai:1999pw,Kawai:1998ab}.
These encourage the future exploration of SSR in string-inspired inflation scenarios.

\acknowledgements 
\noindent

The work of A.A.\ is supported by the National Science Foundation of China (NSFC) 
through the grant No.\ 12350410358; 
the Talent Scientific Research Program of College of Physics, Sichuan University, Grant No.\ 1082204112427;
the Fostering Program in Disciplines Possessing Novel Features for Natural Science of Sichuan University, Grant No.2020SCUNL209 and 1000 Talent program of Sichuan province 2021.
Y.C.\ acknowledges the National Key R\&D Program of China (2021YFC2203100), CAS Young Interdisciplinary Innovation Team (JCTD-2022-20), NSFC (12261131497), 111 Project for ``Observational and Theoretical Research on Dark Matter and Dark Energy'' (B23042), Fundamental Research Funds for Central Universities, CSC Innovation Talent Funds, USTC Fellowship for International Cooperation, USTC Research Funds of the Double First-Class Initiative, CAS project for young scientists in basic research (YSBR-006).

\providecommand{\href}[2]{#2}\begingroup\raggedright\endgroup

\end{document}